\documentclass{IOS-Book-Article}

\usepackage{mathptmx}
\usepackage{comment}
\usepackage{amsmath}
\usepackage{plain}
\usepackage{url}
\usepackage{hyperref}
\usepackage{tikz}
\usepackage{subcaption}

\usepackage{pgfplots}
\pgfplotsset{compat=1.18}

\usetikzlibrary{shapes.geometric, positioning, arrows.meta, calc}

\tikzstyle{attr} = [rectangle, rounded corners, minimum width=2.8cm, minimum height=1cm, text centered, draw=black, fill=blue!20]
\tikzstyle{process} = [rectangle, minimum width=3cm, minimum height=1cm, text centered, draw=black, fill=orange!30]
\tikzstyle{decision} = [diamond, draw, fill=red!30, text centered, aspect=2, minimum height=1.2cm]
\tikzstyle{arrow} = [thick, ->, >=stealth]
\begin{document}
\begin{frontmatter}              % The preamble begins here.

%\pretitle{Pretitle}
\title{Geographic Bias and Diversity in AI Evaluation}
%\runningtitle{Geographic Diversity is the Shield Against Bias}

\author[A]{\fnms{Zilong} \snm{Liu}},%
%\thanks{Corresponding Author: Book Production Manager, IOS Press, Nieuwe Hemweg 6B,
%1013 BG Amsterdam, The Netherlands; E-mail:
%bookproduction@iospress.nl.}},
\author[A]{\fnms{Krzysztof} \snm{Janowicz}},
\author[B]{\fnms{Gengchen} \snm{Mai}},
\author[C]{\fnms{Song} \snm{Gao}}
and 
\author[D]{\fnms{Rui} \snm{Zhu}}

%\runningauthor{B.P. Manager et al.}
\address[A]{University of Vienna, Austria}
\address[B]{University of Texas at Austin, USA}
\address[C]{University of Wisconsin-Madison, USA}
\address[D]{University of Bristol, UK}

\begin{abstract}
Among the many challenges hindering the responsible development and deployment of AI, arguably none has faced more intense scrutiny than bias in its various forms. This underscores the widespread concerns across AI researchers that model outputs, e.g., from generative AI, may encode structural distributional imbalances (stemming from training data or model design) that may amplify social inequality or introduce systemic distortions across application domains ranging from biodiversity to disaster mitigation. Yet, relatively little work has investigated the geographical nature of bias or developed measurable benchmarks for what it means for (generative) AI to be unbiased. In this chapter, we investigate this issue through a literature review. As foundation models are reshaping the landscape of bias research, we examine work spanning both the pre-generative AI and generative AI periods. First, we identify a range of geographic biases. These biases span from representation bias in the training data and regional disparities in the factual recall of language models to the tendency of generative AI to over-proportionally favor prototypical places (called defaults). Then, we showcase how recent studies address the latter bias by evaluating geographic diversity in the outputs of generative AI across various cognitive levels, parameter settings, and output modalities.
\end{abstract}

\end{frontmatter}

\thispagestyle{empty}
\pagestyle{empty}

\section{Introduction}
In the pre-generative AI era, AI bias was often framed as a training-data problem or an inferential problem at the level of embedding space, i.e., an abstract, high-dimensional vector space where machines project their learned concepts. For instance, if a model is trained predominantly with male-centric data, it tends to associate males with other concepts that are closer in the embedding space. As a result, the model performs poorly on tasks requiring balanced gender representations. A well-known illustration of this problem is the 300-dimensional word2vec~\cite{mikolov2013efficient,mikolov2013distributed} model trained on Google News articles. Introduced in 2013, word2vec is one of the earliest large-scale neural language models designed to capture the meaning of words based on their surrounding context. Despite its strengths at the time, it was shown to produce relations, such as
$\vec{\text{man}} - \vec{\text{woman}} \approx \vec{\text{computer programmer}} - \vec{\text{homemaker}}$, in its analogical reasoning~\cite{bolukbasi2016man}.

Today, AI researchers are paying increasing attention to bias in model outputs because foundation models~\cite{bommasani2021opportunities} have become the backbone of AI with unprecedented generative capabilities. Foundation models are pre-trained on massive amounts of data, requiring minimal effort in post-training. It is estimated that GPT-3.5, the foundation model underlying an early release of ChatGPT in 2022, was trained on 570 GB of data. Such very large and deep models often only need a few labeled examples to adapt themselves to a new task. This is a capability known as few-shot learning~\cite{brown2020language}, which enables systems like ChatGPT to be skilled at natural language generation and reasoning at the same time.

Among other systems, ChatGPT is a prominent example of the paradigm shift in AI bias research. Currently, it has hundreds of millions of weekly users\footnote{\url{https://openai.com/business/guides-and-resources/chatgpt-usage-and-adoption-patterns-at-work}}, meaning that the general public is also joining the workforce of evaluators due to their everyday access to such chatbot-like AI systems. Consequently, this trend has broadened the scope of AI bias to include misrepresentation, stereotyping, and other forms of harm at the societal level~\cite{naik2023social, qadri2023ai}.

For geographic information science  (GIScience), the focus extends beyond whether an AI model favors males over females to broader (geographic) questions, such as whether it prioritizes the Global North over the Global South and whether users feel marginalized when interacting with it. These questions fundamentally relate to how local populations are represented by AI designed to operate on a global scale. Also, many of these questions can be meaningfully framed and examined using geographical terms.

In this book chapter, we therefore investigate whether AI bias shows a geographical dimension. We begin with reviewing a selection of representative studies (e.g., \cite{mehrabi2021survey,moayeri2024worldbench}) on AI bias from the past five years. From this review, we identify AI biases that can be expressed in geographical terms. In addition, we go further by asking what constitutes its opposite, i.e., what it means for an AI to be unbiased. We find that the study of AI bias has recently motivated the operationalization of  geographic diversity as an ethical principle~\cite{liu2026geodiversityb, liu2026geodiversitya, liu2025operationalizing}. It offers a reference point for evaluating a less studied type of bias in generative AI, namely its over-representation of certain places.%Notably, this principle has been made measurable for the evaluation of AI-generated textual content. 

\section{Geographic Bias in Machine Learning}
\label{sec:bias}
The first large-scale survey on bias in modern machine learning was published in 2021. It investigated the interplay between bias and fairness~\cite{mehrabi2021survey}. Interestingly, this survey outlined three areas where bias takes on an explicit regional or spatial dimension. According to these areas, a fair model is supposed to:
\begin{itemize}
    \item perform consistently well across individuals, regardless of the geographic origin of their data;
    \item account for variations in statistical assumptions that arise from different choices of geographic units when drawing conclusions about individuals;
    \item avoid discriminatory outcomes in algorithmic inference based on their protected attributes, including those of a geographical nature.
\end{itemize}

\subsection{Representation Bias in Training Data}
\begin{figure}[htbp]
\centering
\begin{tikzpicture}[scale=0.9, every node/.style={font=\footnotesize},
    region/.style={rectangle, minimum width=3cm, minimum height=2cm, draw, thick, align=center},
    training/.style={region, fill=blue!20},
    usepop/.style={region, fill=red!30},
    model/.style={rectangle, draw, thick, fill=gray!20, minimum width=2.8cm, minimum height=1.2cm, align=center},
    arrow/.style={thick, ->, >=Stealth}]

% Training regions
\node[training] (R1) {Region 1};
\node[training, right=1cm of R1] (R2) {Region 2};
\node[training, right=1cm of R2] (R3) {Region 3};

% Model node
\node[model, below=1.6cm of R2] (Model) {\textbf{Model}};

% Use region (underrepresented)
\node[usepop, below=1.6cm of Model] (R4) {Region 4};

% Labels
\node[font=\bfseries, above=1.5cm of $(R1)!0.5!(R3)$] (TrainLabel) {Development Sample};
\node[font=\bfseries, below=0.2cm of R4] (UseLabel) {Use Population};

% Dashed box around training regions
\draw[dashed, thick, rounded corners]
    ($(R1.north west) + (-0.3,0.3)$) rectangle
    ($(R3.south east) + (0.3,-0.3)$);

% Training arrow from dashed box to model
\draw[arrow, blue!80!black] 
    ($(R2.south) + (0,-0.3)$) -- 
    (Model.north) 
    node[midway, right=0.1cm, font=\scriptsize, blue!80!black] {Training};

% Inference arrow from model to region 4
\draw[arrow, red!80!black] 
    (Model.south) -- 
    (R4.north) 
    node[midway, right=0.1cm, font=\scriptsize, red!80!black] {Deployment};

\end{tikzpicture}
\caption{
Illustration of representation bias with an example of four regions. The development sample includes Regions 1–3 for training the model. The model is then deployed to Region 4, which is the use population not represented in training.}
\label{fig:representation-bias}
\end{figure}
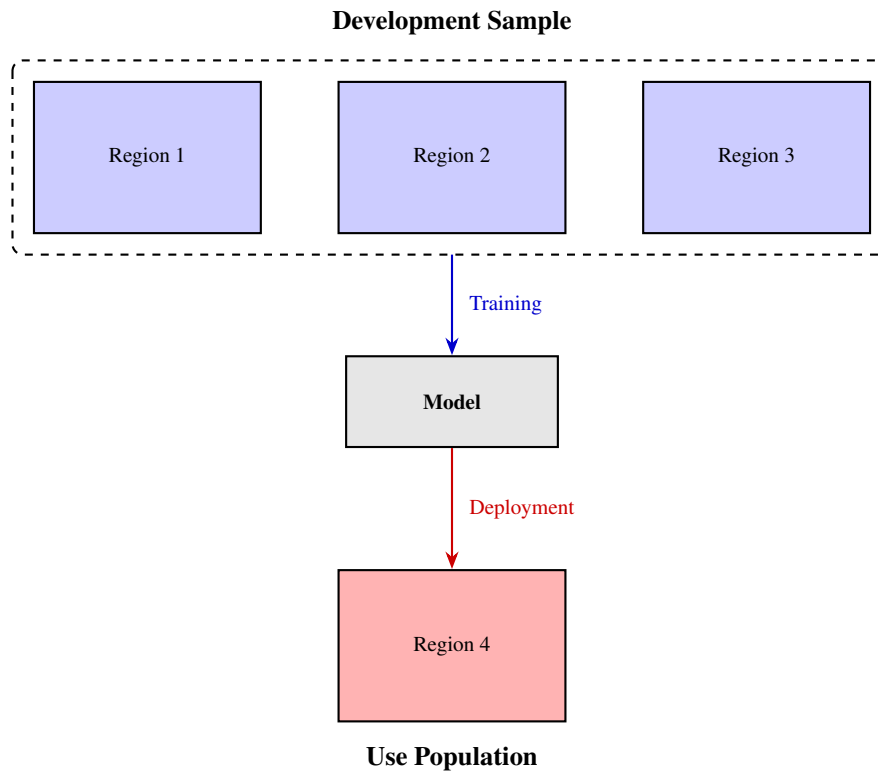

The first area is concerned with the issue of representation bias. According to the survey, it was formalized in a study identifying sources of harm throughout the machine learning lifecycle~\cite{suresh2021framework}. Representation bias arises when the development sample (i.e., a subset of the target population) fails to generalize well for a subset of the use population (i.e., the population a model encounters once deployed); see Figure~\ref{fig:representation-bias}. Within this framework, the lack of geographic diversity was seen as a form of representation bias. This marked the first time that the lack of geographic diversity was prominently identified as a concern for deployment-level modern AI systems, but at this stage, it had not yet been conceptualized as a debiasing standard. Instead, it was merely discussed in the context of constructing benchmark datasets, which are standardized collections of data used to train, validate, and test machine learning models.

Before, Google Brain brought the lack of geographic diversity to attention in their examination of two of the most widely used benchmark datasets in computer vision~\cite{shankar2017no}: Open Images\footnote{\url{https://storage.googleapis.com/openimages/web/index.html}} and ImageNet\footnote{\url{https://www.image-net.org}}. ImageNet, in particular, is significant in advancing AI and computer vision through its annual competition, the ImageNet Large Scale Visual Recognition Challenge~\cite{russakovsky2015imagenet}. In 2012, a major breakthrough occurred when AlexNet~\cite{krizhevsky2012imagenet}, a neural network vision model, won the competition by a wide margin. This success is often regarded as among the earliest success stories of deep learning.

Both Open Images and ImageNet demonstrate a significant lack of geographic diversity despite their wide adoption~\cite{shankar2017no}. At the country level, 32\% of Open Images samples originate from the United States, and 60\% of the data come from just six countries in North America and Europe. In contrast, only 1\% of the samples are from China and 2\% from India, i.e., the two most populous countries in the world. The 2011 fall release of ImageNet shows a similar pattern: 45\% of its images are from the United States, while China and India are represented by just 1\% and 2.1\%, respectively.

What would be the consequences of such a lack of geographic diversity in classic benchmark data used for training? While not systematically evaluated at that time (and even till this date), it was hypothesized that because a benchmark dataset fails to represent a broad and balanced range of geographic regions, it can lead to disparities in model performance across different parts of the world\footnote{Representation bias may also affect the transferability of GeoAI models for between-region inference given the underlying spatial heterogeneity of geographical phenomena~\cite{goodchild2021replication}.}. Some preliminary evidence emerged from the classification behaviors of two vision models~\cite{shankar2017no}, which were pre-trained with Open Images and ImageNet, respectively. First, both models demonstrated a consistent trend of poor performance on two geographically localized stress-test datasets. Specifically, the ImageNet classifier struggled with the object class \texttt{groom, bridegroom}, while the Open Images classifier exhibited similar deficiencies for \texttt{groom} and \texttt{woman}. Second, even within a single stress-test dataset, there were notable disparities in classification performance across countries. For instance, images from the United States and Australia were classified with higher log-likelihoods---indicating that a model has greater confidence in assigning the correct label---compared with those from Ethiopia and Pakistan.
%Sometimes, the ImageNet model misclassified \textit{groom} (or \textit{bridegroom}) images in a stress-test dataset provided by Hyderabad-based crowdsourcing as \textit{chain mail}, \textit{cloth}, \textit{academic gown}, \textit{vestment}, and so forth. 

\subsection{Aggregation Bias in Statistical Inference}
The second area is concerned with the issue of aggregation bias. It was first identified in the same aforementioned study that formalized representation bias~\cite{suresh2021framework}. Aggregation bias arises from using a one-size-fits-all model and applying it to data with substantial differences across sub-groups. It refers to the bias of drawing conclusions about individuals based on generalized assumptions that may only hold true at the population level.

Aggregation bias becomes geographical when it is in the form of the modifiable areal unit problem (MAUP)\footnote{Some geographers have proposed to rename it as the ``OpenShaw Effect''~\cite{goodchild2022openshaw}.}~\cite{openshaw1984modifiable}. Geographers have long studied this issue, recognizing it as a problem that cannot be fully resolved. Instead, it is a problem that should be considered explicitly before performing spatial analysis. Geographers have also added an extra layer of complexity to the understanding of aggregation bias: that it arises not only from the changes in the scale of analysis units, but also from their geometric shape (e.g., due to zoning), even when the scale remains constant. One classic example regarding the zoning impact is gerrymandering, which involves manipulating the boundaries of electoral districts to influence the voting power of opposing political parties~\cite{kruse2024bringing}. Thus, MAUP comprises both scale effects and shape effects.

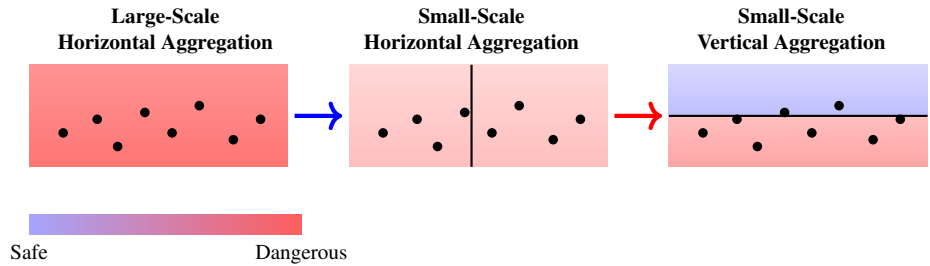
\begin{figure}[htbp]
\centering
\begin{tikzpicture}[scale=0.9, every node/.style={font=\footnotesize}]

% Titles
\node[align=center] at (2,5) {\textbf{Large-Scale} \\ \textbf{Horizontal Aggregation}};
\node[align=center] at (6.5,5) {\textbf{Small-Scale} \\ \textbf{Horizontal Aggregation}};
\node[align=center] at (11.2,5) {\textbf{Small-Scale} \\ \textbf{Vertical Aggregation}};

% 1. Large-Scale Horizontal Aggregation - reddest
\shade[bottom color=red!90, top color=red!70, opacity=0.6] (0,3) rectangle (3.8,4.5);
\foreach \x/\y in {0.5/3.5,1/3.7,1.3/3.3,1.7/3.8,2.1/3.5,2.5/3.9,3/3.4,3.4/3.7} {
    \fill[black] (\x,\y) circle (2pt);
}

% 2. Small-Scale Horizontal Aggregation - lighter red
\shade[bottom color=red!50, top color=red!30, opacity=0.5] (4.7,3) rectangle (6.5,4.5); % left
\shade[bottom color=red!50, top color=red!30, opacity=0.5] (6.5,3) rectangle (8.5,4.5); % right

% Draw boundary line
\draw[black, thick] (6.5,3) -- (6.5,4.5);

\def\xoffset{4.7}
\foreach \x/\y in {0.5/3.5,1.3/3.3,2.1/3.5,3/3.4} {
    \fill[black] ({\x+\xoffset},\y) circle (2pt);
}
\foreach \x/\y in {1/3.7,1.7/3.8,2.5/3.9,3.4/3.7} {
    \fill[black] ({\x+\xoffset},\y) circle (2pt);
}

% 3. Small-Scale Vertical Aggregation - upper blue, bottom red
\shade[bottom color=red!70, top color=red!50, opacity=0.5] (9.4,3) rectangle (13.2,3.75); % bottom red
\shade[bottom color=blue!50, top color=blue!30, opacity=0.5] (9.4,3.75) rectangle (13.2,4.5); % top blue

% Draw explicit boundary line
\draw[black, thick] (9.4,3.75) -- (13.2,3.75);

\def\xoffset{9.4}
\foreach \x/\y in {0.5/3.5, 1/3.7, 1.3/3.3, 1.7/3.8, 2.1/3.5, 2.5/3.9, 3/3.4, 3.4/3.7} {
    \fill[black] ({\x+\xoffset},\y) circle (2pt);
}

% Flow arrows
\draw[->, ultra thick, blue] (3.9,3.75) -- (4.6,3.75);
\draw[->, ultra thick, red] (8.6,3.75) -- (9.3,3.75);

% Horizontal legend below the diagram
\begin{scope}[yshift=2cm]
  % Gradient rectangle
  \shade[left color=blue!50, right color=red!90, opacity=0.7] (0,0) rectangle (4,0.3);
  % Labels
  \node[below] at (0,0) {Safe};
  \node[below] at (4,0) {Dangerous};
\end{scope}

\end{tikzpicture}
\caption{Illustration of aggregation bias using crime incident points, demonstrating both the scale effect (Large-Scale Horizontal Aggregation → Small-Scale Horizontal Aggregation) and the shape effect (Small-Scale Horizontal Aggregation → Small-Scale Vertical Aggregation). Color intensity represents perceived risk: red areas indicate higher danger, and blue areas indicate safer zones.}%Illustration of aggregation bias, demonstrating both the scale effect (Large-Scale Horizontal Aggregation → Small-Scale Horizontal Aggregation) and the shape effect (Small-Scale Horizontal Aggregation → Small-Scale Vertical Aggregation).}
\label{fig:openshaw-effect}
\end{figure}

Figure~\ref{fig:openshaw-effect} illustrates these two dimensions of MAUP. In terms of the scale effect, aggregating all crime observations into two horizontal units smaller than the one large horizontal unit may lead to a different statistical conclusion about regional safety. With respect to the shape effect, repartitioning the same area into two vertical units yet produces another different conclusion.

\subsection{Discrimination Against Protected Attributes}
The third area is concerned with the issue of discrimination against protected attributes of humans in high-stakes decision making. It was emphasized that machine learning models need to reduce their discrimination against protected attributes in such settings.

According to the survey, protected attributes were summarized in relation to two federal laws in the United States in a study on fairness~\cite{chen2019fairness}. The two laws are the Fair Housing Act~\cite{act1968fair} and Equal Credit Opportunity Act~\cite{act2018equal}, which govern housing and credit application, respectively. The shared protected attributes under these laws include national origin, race, religion, sex, and so on. All of these attributes show explicit geographical characteristics in the American context.

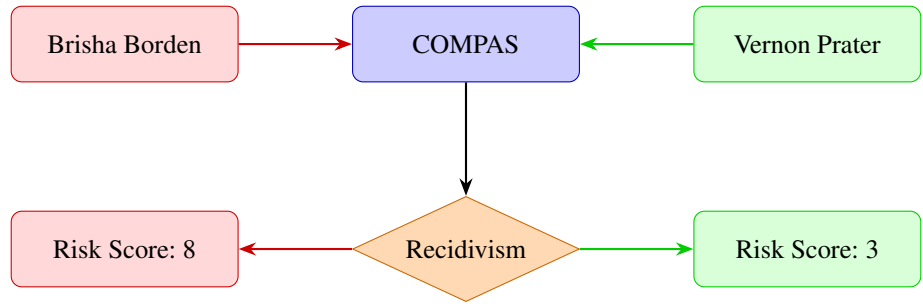
\begin{figure}[htbp]
\centering
\begin{tikzpicture}[
  node distance=2cm and 2.5cm,
  % Left branch styles (red)
  attrLeft/.style={rectangle, draw=red!80!black, rounded corners, fill=red!15, minimum width=3cm, minimum height=1cm, align=center},
  % Right branch styles (green)
  attrRight/.style={rectangle, draw=green!80!black, rounded corners, fill=green!15, minimum width=3cm, minimum height=1cm, align=center},
  % Neutral styles
  process/.style={rectangle, draw=blue!70!black, fill=blue!20, rounded corners, minimum width=3cm, minimum height=1cm, align=center},
  decision/.style={diamond, draw=orange!80!black, fill=orange!30, aspect=2, minimum width=3cm, minimum height=1cm, align=center},
  arrow/.style={->, thick, >=Stealth}
]

% Left branch: African American
\node (origin1) [attrLeft] {Brisha Borden};
\node (model) [process, right=1.5cm of origin1] {COMPAS};
\node (decision) [decision, below=1.5cm of model] {Recidivism};
\node (result1) [attrLeft, left=1.5cm of decision] {Risk Score: 8};

% Right branch: Caucasian American
\node (origin2) [attrRight, right=1.5cm of model] {Vernon Prater};
\node (result2) [attrRight, right=1.5cm of decision] {Risk Score: 3};

% Arrows
\draw[arrow, red!80!black] (origin1) -- (model);
\draw[arrow, green!80!black] (origin2) -- (model);
\draw[arrow] (model) -- (decision);
\draw[arrow, red!80!black] (decision) -- (result1);
\draw[arrow, green!80!black] (decision) -- (result2);

\end{tikzpicture}
\caption{Illustration of how a protected attribute (e.g., race) can influence outcomes in risk-assessment tools such as COMPAS. The left branch (in red) represents the case of Brisha Borden, who received a risk score of 8 despite not reoffending, while the right branch (in green) corresponds to Vernon Prater, who received a risk score of 3 but later reoffended. The two cases originate from the ProPublica story~\cite{angwin2016machine}.}
\label{fig:discrimination}
\end{figure}

The survey frequently referenced Correctional Offender Management Profiling for Alternative Sanctions (COMPAS) as an example of discrimination. In the United States, COMPAS is a tool used to assess the risk of defendants recommitting crimes in the future. ProPublica, a non-profit investigative journalism organization, argued that COMPAS resulted in higher false positive rates for African Americans (see Figure~\ref{fig:discrimination}). These rates were nearly twice as high as those for Caucasian Americans~\cite{angwin2016machine}.% In other words, ProPublica argued that African Americans were more likely to be labeled as high-risk individuals for recidivism, but they did not actually reoffend.

\section{Geographic Bias of Language Models}
The 2021 survey conducted by \cite{mehrabi2021survey} covered bias across various model types, such as vision models and statistical models. Since then, there has been a surge of research moving towards another family of models: language models. 

\subsection{Geographic Disparities in Geoparsing Performance}
Prior research studied language models in their geoparsing performance. Geoparsing consists of two components: toponym recognition \cite{wang2020neurotpr} and toponym resolution \cite{ju2016things}. Toponym recognition is a named entity recognition task focusing on the detection of place names in text, while toponym resolution is the process of mapping those place names to locations. The latter task is often referred to as geocoding or place name disambiguation, as it has to address the challenge of place name ambiguity: that a single place name can refer to multiple locations.

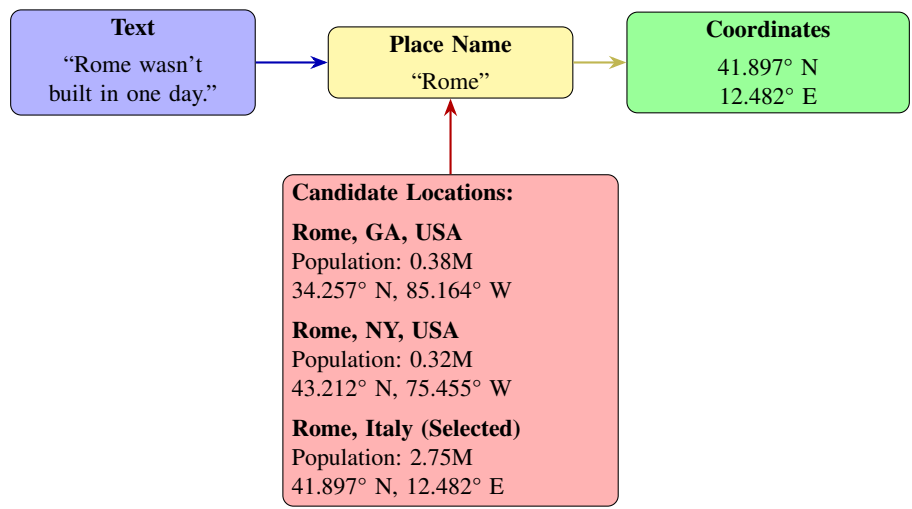
\begin{figure}[htbp]
\centering
\begin{tikzpicture}[node distance=4.2cm, every node/.style={font=\small, rounded corners}, >=Stealth]

    % Nodes
    \node (text) [draw, rectangle, fill=blue!30, align=center, text width=3cm] 
        {\textbf{Text}\\[3pt] ``Rome wasn't built in one day.''};
        
    \node (parse) [draw, rectangle, fill=yellow!40, right of=text, align=center, text width=3cm] 
        {\textbf{Place Name}\\[3pt] ``Rome''};
        
    \node (pop) [draw, rectangle, fill=red!30, below=1cm of parse, align=left, text width=4.2cm] 
        {\textbf{Candidate Locations:}\\[4pt]
         \textbf{Rome, GA, USA} \\ Population: 0.38M \\ 34.257° N, 85.164° W \\[4pt]
         \textbf{Rome, NY, USA} \\ Population: 0.32M \\ 43.212° N, 75.455° W \\[4pt]
         \textbf{Rome, Italy (Selected)} \\ Population: 2.75M \\ 41.897° N, 12.482° E};

    \node (coord) [draw, rectangle, fill=green!40, right of=parse, align=center, text width=3.5cm] 
        {\textbf{Coordinates}\\[3pt] 41.897° N\\ 12.482° E};

    % Arrows
    \draw[->, thick, blue!70!black] (text) -- (parse);
    \draw[->, thick, yellow!70!black] (parse) -- (coord);
    \draw[->, thick, red!70!black] (pop) -- (parse);

\end{tikzpicture}
\caption{A geoparsing pipeline showing how ``Rome'' is extracted and resolved. The final choice reflects an algorithmic bias towards more populous cities.}
\label{fig:geoparsing-bias}
\end{figure}

For both tasks, language models---whether neural or not---struggle to perform equally well across different places~\cite{liu2022geoparsing}\footnote{In fact, there is now a growing body of generative AI research on this problem, expanding the research focus beyond geoparsing alone. For example, generative AI models were found to systematically exhibit errors in their predictions across both objective and subjective topics, affecting regions with lower socioeconomic conditions~\cite{manvi2024large}.}. However, this phenomenon was not attributed solely to representation bias in training data. Instead, it was seen as a potential consequence of multiple interacting sources of bias, including representation bias, aggregation bias, and algorithmic bias; for example, geocoding algorithms often favor places with larger areas or populations in place name disambiguation (see Figure~\ref{fig:geoparsing-bias}).

\subsection{Geographic Disparities in LLM Factual Recall}
Considering the complexity of contributing factors observed in the evaluation of geoparsing performance, an increasing number of studies have begun to focus on the outputs of language models, particularly large language models (LLMs) with millions of parameters. Because LLMs can memorize portions of their training data, evaluation efforts have been undertaken to assess the amount of factual knowledge these models can \textit{recall}.

\begin{figure}[htbp]
\centering
\begin{tikzpicture}[
    node distance=4.2cm,
    every node/.style={font=\small, rounded corners},
    boxInput/.style={draw=blue!70!black, fill=blue!20, thick, rectangle, align=center, text width=2.8cm},
    boxModel/.style={draw=orange!85!black, fill=orange!25, thick, rectangle, align=center, text width=2.5cm},
    boxOutput/.style={draw=green!70!black, fill=green!20, thick, rectangle, align=center, text width=2.8cm},
    arrow/.style={->, thick, >=Stealth}
]

% Nodes with colored boxes
\node (input) [boxInput] 
    {\textbf{Cloze Sentence}\\ ``Paris is the capital of [MASK].''};

\node (model) [boxModel, right of=input]
    {\textbf{LLM}\\ (e.g., BERT)};

\node (output) [boxOutput, right of=model]
    {\textbf{Predicted Token}\\ ``France''};

% Arrows
\draw[arrow, blue!80!black] (input) -- (model);
\draw[arrow, green!80!black] (model) -- (output);

\end{tikzpicture}
\caption{A knowledge-probing pipeline where an LLM predicts a masked token (e.g., ``France'') from a cloze sentence.}
\label{fig:probing}
\end{figure}
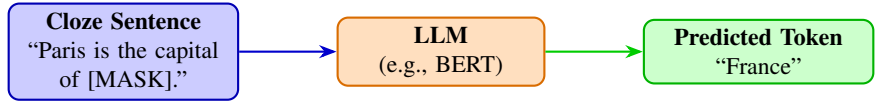

Around 2019, pre-generative AI research initially relied on a task named masked language modeling (which is sometimes referred to as \textit{fill-mask}\footnote{\url{https://huggingface.co/tasks/fill-mask}}), a task that evaluates the ability of a model to recover masked tokens within a cloze sentence~\cite{petroni2019language}. For instance, given a prompt ``Paris is the capital of [MASK]'', an LLM is expected to return ``France'' as the token with the highest log-likelihood (see Figure~\ref{fig:probing}). Generative AI research has expanded benchmarking to include not only masked LLMs but also autoregressive LLMs with a generative design. Detailed instructions are often used to leverage these LLMs' ability to follow human instructions. In this setting, a prompt can take the form of a question, such as ``What country is Paris the capital of'', and an autoregressive LLM can generate a response.

In such a context, geographic bias is regarded as systematic geographical disparities in LLM factual recall \cite{manvi2024geollm,mai2024opportunities}. Using World Bank data, it has been discovered that the error rates in the factual recall of 20 LLMs were 1.5 times higher for Sub-Saharan African countries than for North American countries~\cite{moayeri2024worldbench}.

\section{Geographic Diversity for AI Bias Evaluation}
%Post-ChatGPT research on geographic bias has become increasingly prevalent, yet little work has examined what \textit{unbiased} actually means. Within this framework, geographic diversity has emerged as a natural indicator of an unbiased, expected outcome. Note that it has been mentioned alongside representation bias; see Section~\ref{sec:bias}.

\begin{figure}[htbp]
\centering

% Subfigure (a) - top
\begin{subfigure}[t]{0.8\textwidth}
\centering
\begin{tikzpicture}[
    node distance=4cm,
    every node/.style={font=\small, rounded corners},
    boxInput/.style={draw=blue!70!black, fill=blue!20, thick, rectangle, align=center, text width=3.2cm},
    boxModel/.style={draw=orange!85!black, fill=orange!25, thick, rectangle, align=center, text width=3.2cm},
    boxOutput/.style={draw=green!70!black, fill=green!20, thick, rectangle, align=center, text width=3.2cm},
    arrow/.style={->, thick, >=Stealth}
]

\node (input) [boxInput]
{
\textbf{User Prompt}\\[2mm]
``Describe people in a specific country.''
};

\node (model) [boxModel, right of=input]
{
\textbf{Autoregressive LLM}\\ (e.g., GPT-3.5 Turbo)
};

\node (output) [boxOutput, right of=model]
{
\textbf{Generated Output}\\[1mm]
\begin{tabular}{l}
``Japan''\\
``Japan''\\
``Japan''\\
``Italy''\\
``Italy''\\
\vdots
\end{tabular}
};

\draw[arrow, blue!80!black] (input) -- (model);
\draw[arrow, green!80!black] (model) -- (output);

\end{tikzpicture}
\caption{Experiment adapted from the work of \cite{liu2025operationalizing}.}
\end{subfigure}

% Subfigure (b) - middle
\begin{subfigure}[t]{0.8\textwidth}
\centering
\begin{tikzpicture}[
    node distance=4cm,
    every node/.style={font=\small, rounded corners},
    boxInput/.style={draw=blue!70!black, fill=blue!20, thick, rectangle, align=center, text width=3.2cm},
    boxModel/.style={draw=orange!85!black, fill=orange!25, thick, rectangle, align=center, text width=3.2cm},
    boxOutput/.style={draw=green!70!black, fill=green!20, thick, rectangle, align=center, text width=3.2cm},
    arrow/.style={->, thick, >=Stealth}
]

\node (input) [boxInput]
{
\textbf{User Prompt}\\[2mm]
``Name one country.''
};

\node (model) [boxModel, right of=input]
{
\textbf{Autoregressive LLM}\\ (e.g., ChatGPT-4o)
};

\node (output) [boxOutput, right of=model]
{
\textbf{Generated Output}\\[1mm]
\begin{tabular}{l}
``Canada''\\
``Canada''\\
``Canada''\\
``Canada''\\
``Canada''\\
\vdots
\end{tabular}
};

\draw[arrow, blue!80!black] (input) -- (model);
\draw[arrow, green!80!black] (model) -- (output);

\end{tikzpicture}
\caption{Experiment adapted from the work of \cite{liu2026geodiversitya}.}
\end{subfigure}

% Subfigure (c) - new bottom
\begin{subfigure}[t]{0.8\textwidth}
\centering
\begin{tikzpicture}[
    node distance=4cm,
    every node/.style={font=\small, rounded corners},
    boxInput/.style={draw=blue!70!black, fill=blue!20, thick, rectangle, align=center, text width=3.2cm},
    boxModel/.style={draw=orange!85!black, fill=orange!25, thick, rectangle, align=center, text width=3.2cm},
    boxOutput/.style={draw=green!70!black, fill=green!20, thick, rectangle, align=center, text width=3.2cm},
    arrow/.style={->, thick, >=Stealth}
]

\node (input) [boxInput]
{
\textbf{User Prompt}\\[2mm]
``Generate an image of Vienna.''
};

\node (model) [boxModel, right of=input]
{
\textbf{Autoregressive LLM +  T2I Model}\\ (e.g., GPT-Image-1 + GPT-4o)
};

\node (output) [boxOutput, right of=model]
{
\textbf{Generated Output}\\[1mm]
\begin{tabular}{l}
``Vienna State Opera''\\
``St.\ Stephen's \\Cathedral''\\
``St.\ Stephen's \\Cathedral''\\
``St.\ Stephen's \\Cathedral''\\
``St.\ Stephen's \\Cathedral''\\
\vdots
\end{tabular}
};

\draw[arrow, blue!80!black] (input) -- (model);
\draw[arrow, green!80!black] (model) -- (output);

\end{tikzpicture}
\caption{Experiment adapted from the work of \cite{liu2026geodiversityb}.}
\end{subfigure}

\caption{Experiments in which a user prompts a generative AI model in multiple sessions, producing uneven distributions of place name outputs.}
\label{fig:llm_geobias}
\end{figure}
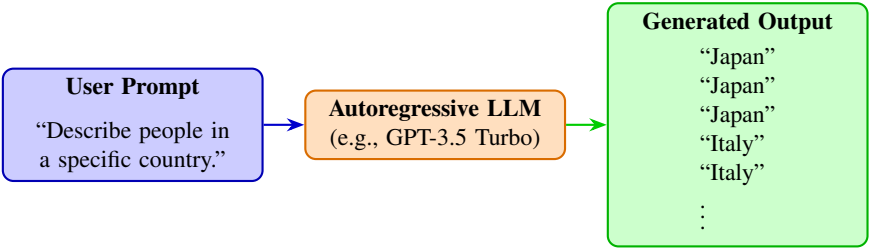
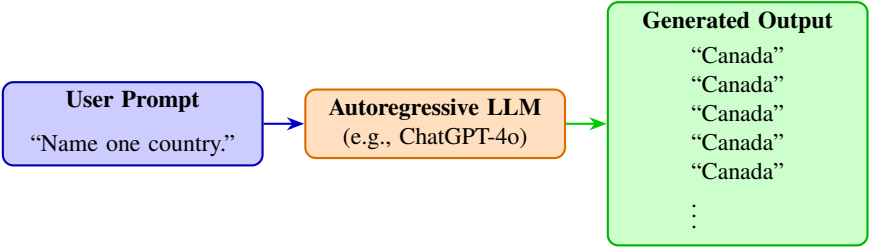
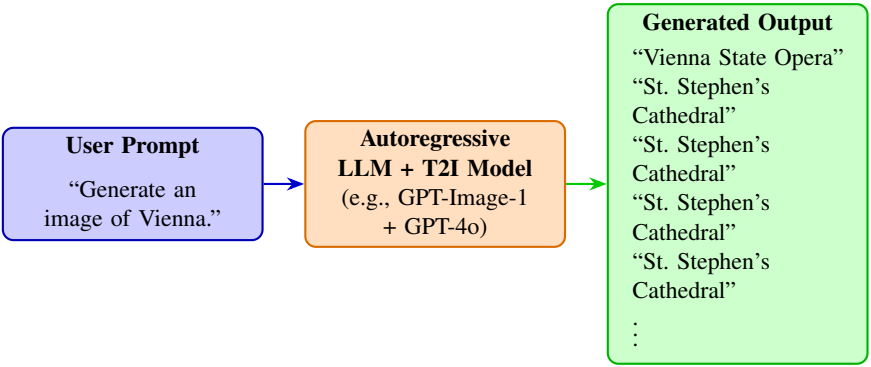

In the generative AI era, it has also been observed that generative AI models tend to exhibit prototypical places across different cognitive levels, ranging from describing the intension of a geographic concept, to identifying instances of a geographic category, and to representing a specific geographic entity (see Figure~\ref{fig:llm_geobias}). For instance, GPT-3.5 Turbo favors Japan when asked to describe people in a specific country~\cite{liu2025operationalizing}. Similarly, ChatGPT-4o shows a preference for Canada when asked to name a country~\cite{liu2026geodiversitya}. In multimodal settings, an image generation pipeline (e.g., GPT-Image-1 and GPT-4o) is biased towards St. Stephan's Cathedral when picturing Vienna~\cite{liu2026geodiversityb}. Collectively, these patterns reveal a new type of geographic bias in AI outputs: a tendency to produce a narrow and skewed set of defaults.

\subsection{Measurement of Geographic Diversity}
Diversity is a natural indicator for quantifying this new geographic bias. This is because it has long been used to quantify similar phenomena on the richness and evenness of species in ecological studies, forming a natural analogy with places in generative AI evaluation. Accordingly, measures of geographic diversity~\cite{liu2025operationalizing} have been proposed to align with well-established measures of species diversity known as the Hill number~\cite{hill1973diversity}. Equation~\ref{eq:geodiversity} defines a general measure of geographic diversity. The parameters are defined as follows.

\begin{itemize}
\item $q$: the order of diversity, controlling the sensitivity of the measure to common versus rare places
\item $S$: the total number of distinct places
\item $p_i$: the proportion of occurrences of the $i^{\text{th}}$ place
\end{itemize}

\begin{equation}
\label{eq:geodiversity}
{}^qD_{\text{geo}} = \left( \sum_{i=1}^{S} p_i^q \right)^{\frac{1}{1-q}}
\end{equation}
This general measure has three commonly used variants corresponding to orders 0, 1, and 2. When $q=0$, the general measure reduces to Equation~\ref{eq:georichness}. It is equivalent to the distinct number of places, thus capturing only their richness (i.e., the cardinality of the set). When $q=1$, the general measure reduces to Equation~\ref{eq:geoshannon}. It is equivalent to the exponential of the Shannon entropy\footnote{Ecologists have highlighted that entropy is not equivalent to diversity~\cite{jost2006entropy}.}~\cite{shannon2001mathematical} (using the natural logarithm), balancing both the richness and evenness of output places. When $q=2$, the general measure reduces to Equation~\ref{eq:geosimpson}. It is equivalent to the Inverse Simpson Index~\cite{simpson1949measurement}, placing greater emphasis on the evenness of output places.

\begin{equation}
\label{eq:georichness}
    {\textstyle {}^0D_{geo}}=S
\end{equation}

\begin{equation}
\label{eq:geoshannon}
    {\textstyle {}^1D_{geo}} = \exp(-\sum_{i=1}^{S} p_i \ln p_i)
\end{equation}

\begin{equation}
\label{eq:geosimpson}
    {\textstyle {}^2D_{geo}} = \frac{1}{\sum_{i=1}^{S} p_i^2}
\end{equation}

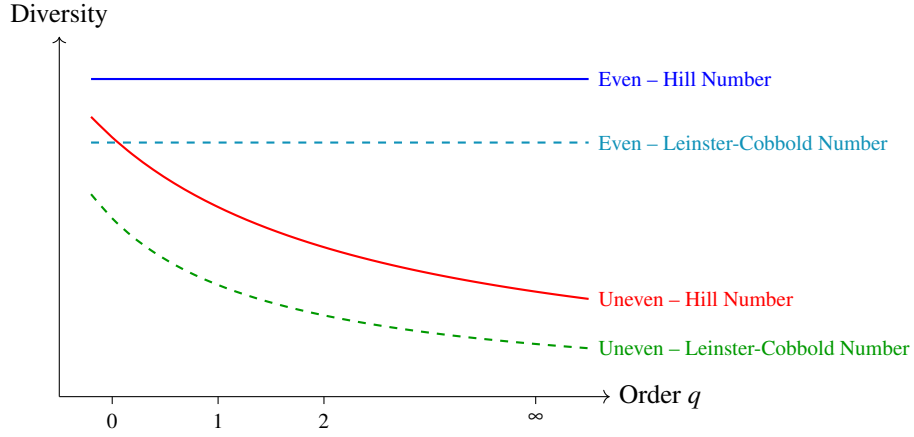
\begin{figure}[htbp]
\centering
\begin{tikzpicture}[scale=1.4]

% Axes
\draw[->] (0,0) -- (5.2,0) node[right] {Order $q$};
\draw[->] (0,0) -- (0,3.4) node[above] {Diversity};

% -------------------------
% EVEN DISTRIBUTIONS
% -------------------------

% Even Hill
\draw[thick,blue] (0.3,3.0) -- (5,3.0)
node[right] {\textcolor{blue}{\footnotesize Even -- Hill Number}};

% Even similarity-sensitive
\draw[thick,cyan!70!black,dashed] (0.3,2.4) -- (5,2.4)
node[right] {\textcolor{cyan!70!black}{\footnotesize Even -- Leinster-Cobbold Number}};

% -------------------------
% UNEVEN DISTRIBUTIONS
% -------------------------

% Uneven Hill
\draw[thick,red] plot[smooth,domain=0.3:5]
(\x,{3/(1+0.45*\x)})
node[right] {\textcolor{red}{\footnotesize Uneven -- Hill Number}};

% Uneven similarity-sensitive
\draw[thick,green!60!black,dashed] plot[smooth,domain=0.3:5]
(\x,{2.4/(1+0.85*\x)})
node[right] {\textcolor{green!60!black}{\footnotesize Uneven -- Leinster-Cobbold Number}};

% q ticks
\foreach \x/\label in {0.5/0,1.5/1,2.5/2,4.5/$\infty$}
  \draw (\x,0) -- (\x,-0.07) node[below] {\footnotesize \label};

\end{tikzpicture}

\caption{Diversity profiles for distributions with and without considering similarity. This plot also shows that the diversity of an even distribution remains constant across orders $q$, while the diversity of an uneven distribution declines as $q$ increases.}
\label{fig:full_diversity_profiles}
\end{figure}

Intuitively, the diversity of observed outputs would decrease as their similarity increases. Therefore, a similarity-sensitive extension to the general measure has also been proposed~\cite{liu2026geodiversityb}, called the Leinster-Cobbold number. This measure also originates from ecological studies~\cite{leinster2012measuring}. Equation~\ref{eq:simdiv} defines the general measure of similarity-sensitive geographic diversity. Here, $\mathbf{Z}$ stands for an $S \times S$ similarity matrix, where $z_{ij} \in [0,1]$ quantifies the similarity between the $i^{\text{th}}$ and $j^{\text{th}}$ output places, and $p_j$ refers to the proportion of occurrences of the $j^{\text{th}}$ place. All other variables are consistent with those in Equation~\ref{eq:geodiversity}. The diversity profile~\cite{leinster2012measuring}, which visually depicts the relationship between the order of diversity and its magnitude, showcases that the similarity-sensitive diversity is consistently smaller than the corresponding similarity-insensitive diversity (see Figure~\ref{fig:full_diversity_profiles}).

\begin{equation}
\label{eq:simdiv}
{}^q\!D_{geo}^{\mathbf{Z}} =
\left(
\sum_{i=1}^{S} p_i 
\Bigg( \sum_{j=1}^{S} z_{ij} p_j \Bigg)^{q-1}
\right)^{\frac{1}{1-q}}
\end{equation}

\subsection{Findings about (a Lack of) Geographic Diversity}
The measurement of geographic diversity yields many interesting findings about the outputs of popular generative AI models. Here, we highlight a few key ones.

\begin{figure}[htbp]
\centering

\begin{tikzpicture}[scale=1.4]

% Axes
\draw[->] (0,0) -- (5.3,0) node[right] {Temperature};
\draw[->] (0,0) -- (0,4.2) node[above] {$^1D_{geo}$};

% X ticks
\foreach \x/\lab in {
0/0, 0.5/0.1, 1/0.2, 1.5/0.3, 2/0.4,
2.5/0.5, 3/0.6, 3.5/0.7, 4/0.8, 4.5/0.9, 5/1.0}
\draw (\x,0.05) -- (\x,-0.05) node[below] {\scriptsize \lab};

% Y ticks
\foreach \y in {0,1,2,3,4}
\draw (0.05,\y) -- (-0.05,\y) node[left] {\scriptsize \y};

% -------------------------------------------------
% chatgpt-4o-latest (BLUE)
% -------------------------------------------------
\draw[thick,blue]
plot[smooth] coordinates {
(0,1)(0.5,1)(1,1)(1.5,1)
(2,1.103005926)
(2.5,1.182870543)
(3,1.254789115)
(3.5,1.384145488)
(4,1.499247754)
(4.5,1.693687258)
(5,1.843077511)
};

\foreach \x/\y in {
0/1,0.5/1,1/1,1.5/1,
2/1.103005926,
2.5/1.182870543,
3/1.254789115,
3.5/1.384145488,
4/1.499247754,
4.5/1.693687258,
5/1.843077511}
\fill[blue] (\x,\y) circle (1.4pt);

\node[blue,right] at (5,1.8) {\footnotesize chatgpt-4o-latest};

% -------------------------------------------------
% gpt-3.5-turbo-0125 (RED DASHED)
% -------------------------------------------------
\draw[thick,red]
plot[smooth] coordinates {
(0,1)
(0.5,1)
(1,1.103005926)
(1.5,1.436122873)
(2,1.443289656)
(2.5,1.762066657)
(3,2.076449555)
(3.5,2.382133847)
(4,2.754908654)
(4.5,3.315972486)
(5,3.857133347)
};

\foreach \x/\y in {
0/1,
0.5/1,
1/1.103005926,
1.5/1.436122873,
2/1.443289656,
2.5/1.762066657,
3/2.076449555,
3.5/2.382133847,
4/2.754908654,
4.5/3.315972486,
5/3.857133347}
\fill[red] (\x,\y) circle (1.4pt);

\node[red,right] at (5,3.9) {\footnotesize gpt-3.5-turbo-0125};

% -------------------------------------------------
% gpt-4-turbo-2024-04-09 (GREEN DOTTED)
% -------------------------------------------------
\draw[thick,green!60!black]
plot[smooth] coordinates {
(0,1)
(0.5,1)
(1,1)
(1.5,1)
(2,1.103005926)
(2.5,1.103005926)
(3,1.30363742)
(3.5,1.522129022)
(4,1.602225318)
(4.5,1.792745568)
(5,1.93327338)
};

\foreach \x/\y in {
0/1,
0.5/1,
1/1,
1.5/1,
2/1.103005926,
2.5/1.103005926,
3/1.30363742,
3.5/1.522129022,
4/1.602225318,
4.5/1.792745568,
5/1.93327338}
\fill[green!60!black] (\x,\y) circle (1.4pt);

\node[green!60!black,right] at (5,2.05) {\footnotesize gpt-4-turbo-2024-04-09};

\end{tikzpicture}

\caption{The average order-1 geographic diversity versus sampling temperature across three studied models from the work of \cite{liu2026geodiversitya}.}
\label{fig:diversity_comparison}
\end{figure}
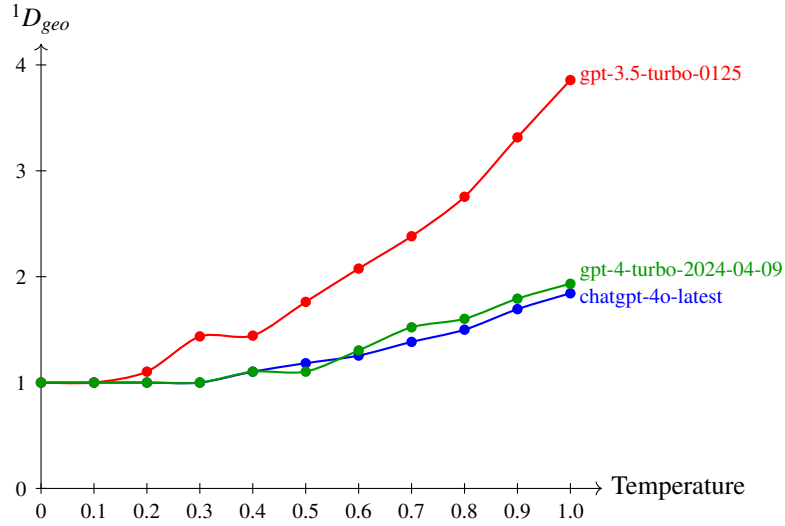

First, geographic diversity becomes smaller when similarity is incorporated as a weighting factor. This is observed for the identified Viennese landmarks in image generation across 6 models~\cite{liu2026geodiversityb}. In practice, this means that generative AI tends to output not only the same prototypical place but also similar types of places. Second, geographic diversity typically increases monotonically with sampling temperature~\cite{ackley1985learning}, which controls the randomness of LLM outputs. This trend is evident for 297 geographic categories collectively across 10 models~\cite{liu2026geodiversitya} (see Figure~\ref{fig:diversity_comparison}). It further suggests that higher geographic diversity may be achieved by adjusting sampling temperature as a model parameter (potentially at the cost of accuracy). Third, newer models do not necessarily exhibit greater geographic diversity than older ones. This is a pattern that can be observed across recent studies on geographic diversity, including comparisons between GPT-3.5 Turbo and later GPT models (see Figure~\ref{fig:diversity_comparison})~\cite{liu2026geodiversitya}. As such, current pre-training and post-training efforts are likely to produce less geographically diverse models.

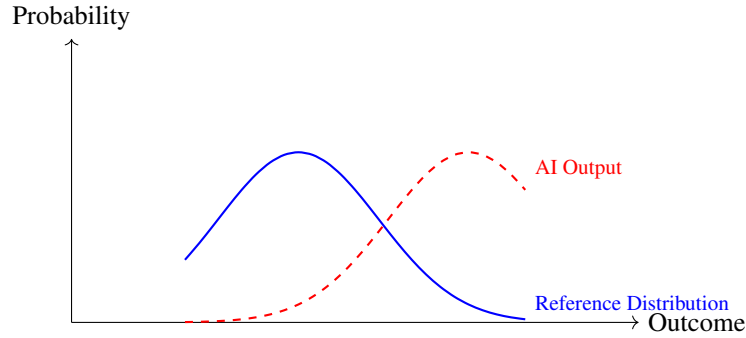
\begin{figure}[htbp]
\centering

% Reference Distribution plot
\begin{tikzpicture}[scale=1.5]

% Axes
\draw[->] (0,0) -- (5,0) node[right] {Outcome};
\draw[->] (0,0) -- (0,2.5) node[above] {Probability};

% Population distribution (normal curve)
\draw[thick,blue] plot[smooth,domain=1:4] 
  (\x, {1.5*exp(-(\x-2)^2)})
  node[right, yshift=6pt] {\textcolor{blue}{\footnotesize Reference Distribution}};

% AI output distribution shifted (biased)
\draw[thick,red,dashed] plot[smooth,domain=1:4] 
  (\x, {1.5*exp(-(\x-3.5)^2)})
  node[right, yshift=8pt] {\textcolor{red}{\footnotesize AI Output}};

\end{tikzpicture}

\caption{Bias as a systematic deviation (with respect to the mean in this example) of the AI output from a statistical reference distribution.}
\label{fig:bias}
\end{figure}

\section{Conclusions}
The introduction of geographic diversity highlights that diversity should not only be articulated but also be made spatially explicit and measurable for the evaluation of (generative) AI systems. This leads to two subsequent research questions. First, what should be the expected geographic diversity? From a GIScience perspective, measures of geographic diversity should not be solely used to benchmark (Gen)AI models; they should also serve to model the \textit{ground-truth} diversity present in \textit{reality}---much in the same way that Moran’s I~\cite{moran1950notes} has become a classical measure of spatial dependence~\cite{anselin1989special} by quantifying the degree of spatial clustering in the attributes of environmental observations. Second, if geographic diversity can be measured, can geographic bias be measured as well? While the definition of bias differs across the machine learning lifecycle, it can be probabilistically modeled as a systematic deviation from a reference distribution (see Figure~\ref{fig:bias}) \cite{wu2024torchspatial}. As long as human expectations, goals, or societal norms can be represented by this distribution, those biases---such as representation bias when understood as a lack of geographic diversity \textit{relative to reality}---can be quantified, thereby contributing to the ongoing debate about AI (geo-)alignment~\cite{janowicz2025whose,sorensen2024roadmap}.

\bibliographystyle{plainurl}
\bibliography{reference}

\end{document}